\documentclass[sigconf, nonacm]{acmart}

\usepackage{tikz}
\usetikzlibrary{positioning, arrows.meta, calc, patterns, decorations.pathreplacing}
\usepackage{pgfplots}
\usepackage{balance}
\pgfplotsset{compat=1.18}

\newcommand\vldbdoi{XX.XX/XXX.XX}
\newcommand\vldbpages{XXX-XXX}
\newcommand\vldbvolume{19}
\newcommand\vldbissue{1}
\newcommand\vldbyear{2026}
\newcommand\vldbauthors{\authors}
\newcommand\vldbtitle{\shorttitle} 
\newcommand\vldbavailabilityurl{https://github.com/redpanda-data/ring_shuffle}
\newcommand\vldbpagestyle{plain} 

\begin{document}
\title{One Ring to Shuffle Them All: Scalable Intra-Process Data Redistribution with Ring-Buffer Shuffle in Redpanda Oxla}

\author{Adam Szymański}
\affiliation{%
  \institution{Redpanda}
  \city{Warsaw}
  \state{Poland}
}
\email{adam.szymanski@redpanda.com}

\author{Tyler Akidau}
\affiliation{%
  \institution{Redpanda}
  \city{Seattle}
  \state{USA}
}
\email{takidau@redpanda.com}


\begin{abstract}
As server CPUs scale to dozens and now hundreds of cores per socket, parallel query engines must rethink how they redistribute data between threads. 
Partitioned operators such as hash joins and aggregations require frequent data redistribution across threads, yet existing intra-process shuffle designs fundamentally fail to scale with core count: \textit{batch partitioning} avoids cross-thread synchronization in the hot path but materializes all intermediate data, introduces a global producer/consumer barrier, and requires a consumption approach with low cache locality, while \textit{channel-based streaming} avoids materialization but incurs per-channel synchronization that scales poorly with core count.
As core counts rise, these architectural tradeoffs increasingly prevent engines from fully utilizing modern hardware.

We present a \textit{ring-buffer streaming} shuffle design that addresses these shortcomings through lock-free atomic slot acquisition into fixed-size batch groups, achieving amortized $O(1)$ synchronization cost per batch and $O(M)$ memory independent of input size. Ring-buffer shuffle has been implemented in Redpanda's Oxla query engine for two years, where it currently powers production queries for Redpanda SQL users.

We evaluate all three approaches on a 72-core NVIDIA GraceHopper, a 192-core dual-socket AWS Graviton4, and a 96-core (192-thread) AMD EPYC. On a 72-core single-socket system the ring buffer outperforms channel streaming by up to 44\% and batch partitioning by up to 79\%; at 192 cores the advantage over channel grows to over 100\% and over 300\% versus batch partitioning.  Even so, on chiplet architectures with many partitioned L3 caches, the shared atomic counter becomes a cross-die bottleneck and channel-based streaming remains competitive.

End-to-end Graviton4 evaluation on TPC-H (21 queries) and ClickBench (43 queries) shows the advantage is workload-shape-dependent: the ring buffer wins TPC-H in aggregate (1.07$\times$) across diverse query shapes, but a handful of consumer-heavy patterns---\texttt{COUNT(DISTINCT)}, wide aggregates---favor channels and swing ClickBench's aggregate the other way.
\end{abstract}

\maketitle

\pagestyle{\vldbpagestyle}
\begingroup\small\noindent\raggedright\textbf{PVLDB Reference Format:}\\
\vldbauthors. \vldbtitle. PVLDB, \vldbvolume(\vldbissue): \vldbpages, \vldbyear.\\
\href{https://doi.org/\vldbdoi}{doi:\vldbdoi}
\endgroup
\begingroup
\renewcommand\thefootnote{}\footnote{\noindent
This work is licensed under the Creative Commons BY-NC-ND 4.0 International License. Visit \url{https://creativecommons.org/licenses/by-nc-nd/4.0/} to view a copy of this license. For any use beyond those covered by this license, obtain permission by emailing \href{mailto:info@vldb.org}{info@vldb.org}. Copyright is held by the owner/author(s). Publication rights licensed to the VLDB Endowment. \\
\raggedright Proceedings of the VLDB Endowment, Vol. \vldbvolume, No. \vldbissue\ %
ISSN 2150-8097. \\
\href{https://doi.org/\vldbdoi}{doi:\vldbdoi} \\
}\addtocounter{footnote}{-1}\endgroup

\ifdefempty{\vldbavailabilityurl}{}{
\vspace{.3cm}
\begingroup\small\noindent\raggedright\textbf{PVLDB Artifact Availability:}\\
The source code, data, and/or other artifacts have been made available at \url{\vldbavailabilityurl}.
\endgroup
}

\section{Introduction}

Modern analytical database systems exploit multi-core hardware by executing query operators in parallel across threads. A wide range of relational operators---hash joins, hash aggregations, window functions---require that at some point during processing, all data related to a given key converges on a single thread. One way to achieve this is through shared synchronized state: multiple threads can concurrently read and update a shared data structure such as a partitioned hash map, avoiding explicit data movement altogether. Today this approach is used for specific operators---notably parallel hash join builds~\cite{Leis14, PostgreSQLParallelHash} and, more recently, group-by aggregation~\cite{XueMarcus25}---and ongoing research suggests it may become competitive in a broader range of scenarios. Nevertheless, many production systems still rely on explicit \emph{data redistribution}: given a stream of data produced in parallel by $M$ threads, route items to $N$ consumer threads based on a partitioning function. Importantly, what gets redistributed need not be the original input rows---systems may shuffle partial aggregates, hash map partitions, or other intermediate representations. This redistribution operation, commonly referred to as \emph{shuffle} or \emph{exchange}, is on the critical path of every partitioned operator and is a fundamental primitive in parallel query execution~\cite{Graefe90, Graefe94}.

The database community has extensively studied shuffle in the distributed setting---network protocols, serialization formats, compression, fault tolerance, and spilling strategies have all received significant attention~\cite{Presto23, Rodiger16}. By contrast, the \emph{intra-process} variant of the problem---exchanging rows between threads within a single process on a single machine---has received far less dedicated treatment. It is often regarded as a solved problem, either subsumed into operator-specific implementations or inherited wholesale from the Volcano exchange model~\cite{Graefe90}. Yet on modern many-core servers with 64 to 256 or more cores, this local redistribution is itself a significant performance bottleneck. As core counts grow, per-core memory bandwidth has not scaled proportionally---making cache misses increasingly costly. With the rapid growth in core counts, the efficiency of data movement between threads has become more important than ever.

Existing approaches to intra-process shuffle face a fundamental tension between \emph{memory overhead} and \emph{contention}. Batch-oriented designs, exemplified by morsel-driven systems such as HyPer and DuckDB~\cite{Leis14, Kuiper25}, have each thread accumulate input data locally, partition it, and then merge matching partitions from all threads after a barrier. This avoids cross-thread synchronization during the accumulation phase but requires full materialization of intermediate results, consuming memory proportional to the input size and causing cache misses when consumers access data written by distant threads. Streaming designs, rooted in the Volcano exchange operator~\cite{Graefe90} and employed by systems such as StarRocks~\cite{StarRocks} and DataFusion~\cite{DataFusion24}, route rows to consumers as they are produced via per-partition channels. This avoids materialization but introduces synchronization on every channel operation; the resulting contention grows with the number of threads and becomes a scalability bottleneck on many-core hardware.

In this paper, we present a novel ring-buffer-based shuffle design that navigates between both extremes. Our approach streams data between producers and consumers without full materialization, yet achieves $O(1)$ memory per core and minimal cross-thread contention through a lock-free protocol based on atomic counters and fixed-size batch groups. We have implemented and deployed this design in Oxla, a distributed analytical SQL engine developed at Redpanda, where it serves as the shuffle primitive for all partitioned operators in production workloads.

This paper makes the following contributions:
\begin{itemize}
  \item We frame intra-process shuffle as a first-class problem, distinct from distributed shuffle, and identify the memory-versus-contention tradeoff that governs existing designs.
  \item We present a lock-free ring-buffer shuffle design with $O(1)$ memory per core, providing temporal data locality while avoiding the contention-scaling problems of channel-based approaches.
  \item We evaluate all three approaches---batch partitioning, channel-based streaming, and ring-buffer streaming---in controlled benchmarks across varying core counts and data distributions.
\end{itemize}

The remainder of this paper is organized as follows: Section \ref{sec:related} discusses related work on parallel query execution and exchange operators. Section \ref{sec:design} describes the three shuffle approaches in detail. Section \ref{sec:evaluation} presents our experimental evaluation. Section \ref{sec:production} discusses production deployment experience in Oxla. Section \ref{sec:conclusion} concludes.

\section{Related Work}
\label{sec:related}

We organize related work along the conceptual progression from the foundational exchange model to the two dominant families of intra-process redistribution (batch partitioning and channel-based streaming), and finally to shared-state alternatives that sidestep explicit data movement.

\subsection{The Exchange Operator}
\label{sec:exchange}

Graefe introduced the \emph{exchange} operator in the Volcano system as a meta-operator that encapsulates parallelism without modifying the logic of data-processing operators~\cite{Graefe90, Graefe94}. Exchange interposes between a producer and a consumer, using bounded packet queues in shared memory to transfer data between forked processes. A support function---hash, range, or round-robin---routes each packet to the appropriate consumer port. Within each process, execution remains demand-driven (iterators pull tuples); between processes, communication is data-driven (producers push packets into queues). This clean separation of parallelism from operator logic made the exchange approach the template for plan-driven parallel query execution, and most modern systems---including Presto~\cite{Presto23}, StarRocks, and DataFusion---still employ exchange-derived operators for data redistribution.

However, the original exchange model was designed for relatively low process counts and does not address the memory hierarchy effects that dominate on modern many-core hardware. R\"{o}diger et al.\ explicitly criticize exchange-based architectures for ``unnecessary materialization'' and inflexibility when scaling to high core counts~\cite{Rodiger16}. The two families of approaches described below can be understood as different responses to these limitations.

\subsection{Batch Partitioning}
\label{sec:batch}

Morsel-driven parallelism, introduced in HyPer~\cite{Leis14}, eliminates the exchange operator entirely for intra-process parallelism. Instead of routing tuples through channels, threads pull fixed-size \emph{morsels} from a shared input, process them independently into thread-local storage, and synchronize only at pipeline boundaries through barrier-like phase transitions. A dispatcher assigns morsels with NUMA-aware scheduling, ensuring threads primarily access local memory. For hash joins, the build phase inserts into a shared hash table via compare-and-swap (CAS); redistribution is implicit in the transition between pipeline phases rather than explicit in a dedicated operator.

Bandle et al.\ integrate radix-partitioned hash joins into the morsel-driven framework in Umbra~\cite{Bandle21}. Their approach uses two-pass radix partitioning with software write-combining buffers (SWWCBs)~\cite{Polychroniou14} and non-temporal stores to avoid polluting the cache hierarchy. Each thread partitions into thread-local pages; after a barrier, partitions from all threads are concatenated by exchanging metadata pointers rather than copying data. The key contribution is a cost model that predicts when radix partitioning pays off relative to a non-partitioned hash join, based on table sizes and available cache. In a companion paper, Bandle and Giceva decompose the exchange operator into composable \emph{sub-operators}---scan, map, scatter, gather, fold---providing a vocabulary for reasoning about the individual steps of radix partitioning~\cite{Bandle21b}.

Kuiper et al.\ extend morsel-driven radix partitioning to external (larger-than-memory) hash joins in DuckDB~\cite{Kuiper25}. Threads partition into spillable pages using a two-phase radix scheme; thread-local partitions are merged after a barrier. The design maintains the morsel-driven property that threads operate independently during partitioning, synchronizing only to merge partition metadata.

The key tradeoff of batch partitioning is clear: no cross-thread synchronization during the accumulation phase, but a barrier is required before consumers can proceed, and the accumulated data must be materialized in its entirety. When thread-local hash tables provide significant reduction---as in low-cardinality aggregation---the materialized state is compact and the merge phase is inexpensive. However, when key cardinality is high or the operation does not aggregate (e.g., a partitioned join build), memory consumption is proportional to the input size, and consumers accessing data written by distant cores incur cache misses during the merge phase. Balkesen et al.\ provide a comprehensive experimental study of these cache effects for multi-core hash joins~\cite{Balkesen13}.

\subsection{Channel-Based Streaming}
\label{sec:channel}

The alternative family preserves the streaming nature of the original exchange model, routing data to consumers as it is produced rather than accumulating it. Modern implementations replace Volcano's process-level packet queues with thread-level channels using contemporary concurrency primitives. StarRocks~\cite{StarRocks} implements a \texttt{LocalExchanger} with multiple strategies---hash partitioning, round-robin, broadcast, and adaptive passthrough---where sink operators push chunks into per-partition buffers and source operators pull from them. DataFusion's~\cite{DataFusion24} \texttt{RepartitionExec} creates $N \times M$ asynchronous channels (for $N$ input and $M$ output partitions) with gate-based backpressure, and supports sort-preserving repartition and spill-to-disk under memory pressure.

At the lowest level, the choice of channel implementation matters significantly. Baumstark and Pohl benchmark lock-free single-producer single-consumer (SPSC) ring buffers against mutex-based queues for inter-thread tuple transfer, demonstrating 4--5$\times$ throughput improvement with lock-free designs~\cite{Baumstark19}. Their work highlights that the synchronization primitive---not just the architectural pattern---is a critical performance factor.

Velox, Meta's unified execution engine, also employs exchange operators for local data redistribution~\cite{Velox22}. While the published description focuses on the broader execution framework rather than exchange internals, the architecture appears to follow the channel-based pattern, and the authors identify shuffle as a performance-sensitive component for partitioned operators.

The fundamental tradeoff of channel-based streaming is the inverse of batch partitioning: it avoids full materialization (achieving low memory overhead and low latency to first results), but requires synchronization on every channel operation. As the number of threads grows, contention on channel data structures becomes a scalability bottleneck.

\subsection{Shared Synchronized State}
\label{sec:shared}

A third approach avoids explicit redistribution entirely by having multiple threads concurrently access a shared data structure. Leis et al.\ use CAS-based insertion into a shared hash table for the join build phase~\cite{Leis14}. PostgreSQL's parallel hash join has multiple workers cooperatively build a shared hash table in dynamic shared memory~\cite{PostgreSQLParallelHash}. More recently, Xue and Marcus show that purpose-built concurrent hash tables can match partitioning approaches for GROUP BY aggregation, arguing that shared state is viable and underexplored as a general strategy~\cite{XueMarcus25}.

These techniques are currently operator-specific---tied to hash joins or aggregations---rather than general-purpose exchange mechanisms. They represent an important complementary direction but are outside the scope of this paper, which focuses on general data redistribution between parallel pipeline stages.

\section{Design}
\label{sec:design}

We now describe three approaches to intra-process shuffle in detail: \textit{batch partitioning}, \textit{channel-based streaming}, and \textit{ring-buffer streaming}. The former two see broad adoption across the industry, as detailed in the previous section, while the latter is our novel approach from Redpanda Oxla.

All three solve the same problem: given $M$ producer threads that generate columnar batches and $N$ consumer threads that must each receive all rows assigned to them by a partitioning function $h$, route every input row to its designated consumer. A \emph{batch} is a fixed-capacity column-oriented container holding up to $B$ rows. Correctness requires that every input row is delivered to exactly one consumer, determined by $h$, with no duplication or loss.

The three designs differ in when and how they synchronize producers and consumers. We evaluate them along four axes: (1)~total memory footprint, (2)~synchronization rate as a function of thread count, (3)~latency to first output, and (4)~consumer cache locality. \autoref{tab:design-comparison} summarizes the key properties of the three approaches.

\begin{table}[t]
\caption{Comparison of shuffle design properties. $M$: producers, $N$: consumers.}
\label{tab:design-comparison}
\small
\setlength{\tabcolsep}{3pt}
\begin{tabular}{lccc}
\toprule
\textbf{Property} & \textbf{Batch} & \textbf{Channel} & \textbf{Ring-Buffer} \\
\midrule
Memory & $O(|input|)$ & $O(N)$ & $O(M)$ \\
Sync rate$^\dagger$ & none$^*$ & $O(M)$ & $O(1)$ \\
Latency & barrier & streaming & streaming \\
Consumer temporal locality & stale & uncorrelated & correlated \\
\bottomrule
\multicolumn{4}{l}{\scriptsize $^*$No sync during accumulation; barrier at end.} \\
\multicolumn{4}{l}{\scriptsize $^\dagger$Total lock/mutex acquisitions per time unit across all producers.}
\end{tabular}
\end{table}

All three approaches share a common preprocessing step: \emph{batch indexing}. When a producer receives an input batch of up to $B$ rows, it evaluates $h$ for every row to determine each row's target partition. It then constructs an index structure that allows any consumer to efficiently extract the rows belonging to its partition. This indexing requires a single pass over the batch ($O(B)$) and is entirely thread-local. What differs between the three designs is what happens after indexing: batch partitioning copies rows into thread-local buffers; channel-based and ring-buffer streaming both push indexed batches onward without copying, but differ in how they coordinate between producers and consumers.

\subsection{Batch Partitioning}
\label{sec:design-batch}

\begin{figure}[t]
\centering
\begin{tikzpicture}[
  >=Stealth,
  producer/.style={draw, minimum width=1.2cm, minimum height=0.6cm, font=\small},
  consumer/.style={draw, minimum width=1.2cm, minimum height=0.6cm, font=\small},
  buf/.style={draw, minimum width=0.4cm, minimum height=0.4cm, font=\scriptsize, inner sep=1pt},
  phase/.style={font=\small\itshape, text=gray},
]
  \node[producer] (p1) at (0, 0) {$P_1$};
  \node[producer] (p2) at (0, -1.0) {$P_2$};
  \node[producer] (p3) at (0, -2.0) {$P_3$};

  \foreach \p/\y in {1/0, 2/-1.0, 3/-2.0} {
    \node[buf, fill=blue!15] (b\p a) at (2.0, \y+0.15) {\tiny $h{=}0$};
    \node[buf, fill=red!15] (b\p b) at (2.0, \y-0.15) {\tiny $h{=}1$};
    \draw[->, thin] (p\p.east) -- (b\p a.west);
    \draw[->, thin] (p\p.east) -- (b\p b.west);
  }

  \draw[dashed, thick, gray] (3.0, 0.6) -- (3.0, -2.6) node[below, phase] {barrier};

  \node[consumer, fill=blue!15] (c1) at (5.0, -0.5) {$C_1$};
  \node[consumer, fill=red!15] (c2) at (5.0, -1.5) {$C_2$};

  \foreach \p in {1, 2, 3} {
    \draw[->, thin, blue!60] (b\p a.east) -- (c1.west);
    \draw[->, thin, red!60] (b\p b.east) -- (c2.west);
  }

  \node[phase] at (1.0, 1.0) {accumulate};
  \node[phase] at (5.0, 1.0) {merge};
\end{tikzpicture}
\caption{Batch partitioning with $M{=}3$ producers and $N{=}2$ consumers. Producers write to thread-local partition buffers without synchronization. After a barrier, each consumer reads its partition from all producers.}
\Description{Diagram showing three producers writing to thread-local partition buffers, a dashed barrier line, and two consumers reading from all buffers.}
\label{fig:batch-design}
\end{figure}
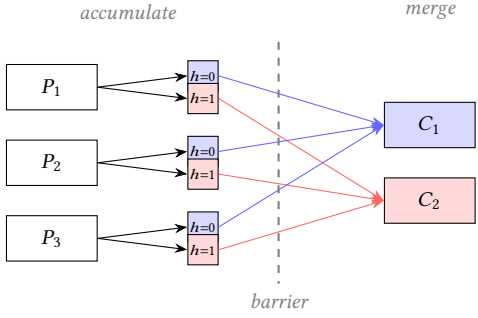

In batch partitioning, each producer accumulates its entire input locally before any consumer begins processing. The approach proceeds in three phases, illustrated in \autoref{fig:batch-design}.

In the \emph{accumulation phase}, each of the $M$ producer threads pulls input morsels and partitions rows by $h$ into $N$ thread-local buffers. Because each thread writes only to its own memory, no cross-thread synchronization is required during this phase.

A \emph{barrier} follows: all producers must complete before any consumer can start. This is a global synchronization point that separates the write phase from the read phase.

In the \emph{merge phase}, each consumer~$j$ reads partition~$j$ from all $M$ producers' thread-local buffers. Since the data was written by remote threads---potentially on distant NUMA nodes---these reads typically miss the local caches and incur main memory latency.

The total memory consumed is $O(|input|)$: the entire input is materialized in the thread-local buffers before consumers begin. When thread-local aggregation provides significant reduction (e.g., low-cardinality GROUP~BY), the materialized state is only $O(M \times K)$ where $K$ is the number of distinct keys, and the merge phase is correspondingly cheaper. However, for operations that do not aggregate---such as a partitioned join build with high key cardinality---the full input must be materialized, and the barrier prevents any pipelining between production and consumption.

Beyond the latency cost of the barrier itself, the two-phase structure introduces a \emph{tail-latency} problem: within each phase, the available cores are fully utilized only until the fastest thread finishes. Partition skew or uneven morsel sizes cause some threads to complete earlier than others, leaving their cores idle until the slowest thread reaches the barrier. This underutilization occurs twice---once at the end of the accumulation phase and once at the end of the merge phase. In contrast, streaming approaches naturally self-balance: when a producer stalls (e.g., on a full channel), the OS schedules the co-resident consumer, keeping the core occupied.

\subsection{Channel-Based Streaming}
\label{sec:design-channel}

\begin{figure}[t]
\centering
\begin{tikzpicture}[
  >=Stealth,
  producer/.style={draw, minimum width=1.2cm, minimum height=0.6cm, font=\small},
  consumer/.style={draw, minimum width=1.2cm, minimum height=0.6cm, font=\small},
  chan/.style={draw, minimum width=1.4cm, minimum height=0.4cm, font=\scriptsize, rounded corners=2pt},
  phase/.style={font=\small\itshape, text=gray},
]
  \node[producer] (p1) at (0, 0) {$P_1$};
  \node[producer] (p2) at (0, -1.0) {$P_2$};
  \node[producer] (p3) at (0, -2.0) {$P_3$};

  \node[chan, fill=blue!15] (ch1) at (2.5, -0.5) {channel$_1$};
  \node[chan, fill=red!15] (ch2) at (2.5, -1.5) {channel$_2$};

  \foreach \p in {1, 2, 3} {
    \draw[->, thin, blue!60] (p\p.east) -- (ch1.west);
    \draw[->, thin, red!60] (p\p.east) -- (ch2.west);
  }

  \node[consumer, fill=blue!15] (c1) at (5.0, -0.5) {$C_1$};
  \node[consumer, fill=red!15] (c2) at (5.0, -1.5) {$C_2$};

  \draw[->, thick, blue!60] (ch1.east) -- (c1.west);
  \draw[->, thick, red!60] (ch2.east) -- (c2.west);

  \node[phase] at (1.2, 0.8) {push by $h$};
  \node[phase] at (4.0, 0.8) {pull};
\end{tikzpicture}
\caption{Channel-based streaming with $M{=}3$ producers and $N{=}2$ consumers. Each producer routes rows by $h$ and pushes to the corresponding channel. Consumers pull from their dedicated channel. Synchronization occurs on every push and pull.}
\Description{Diagram showing three producers connected to two channels, each channel feeding one consumer.}
\label{fig:channel-design}
\end{figure}
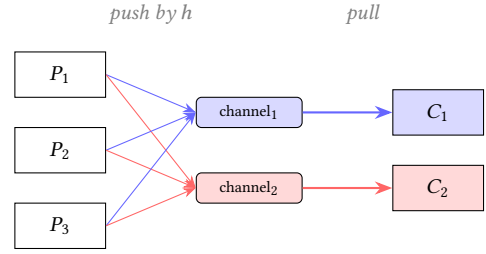

Channel-based streaming preserves the data-driven nature of the Volcano exchange model: producers route batches to consumers as they are produced, with no barrier between production and consumption. The architecture is shown in \autoref{fig:channel-design}.

Each output partition is associated with a \emph{channel}---a bounded buffer with synchronization primitives that supports concurrent push and pull operations. After batch indexing, the producer pushes the indexed batch to each of the $N$ output channels; the consumer on the other end extracts its partition's rows using the precomputed index. Consumers pull from their channels, blocking when no data is available.

Backpressure is inherent: when a channel is full, the producer blocks until the consumer drains space. This bounds total memory at $O(N)$---one channel per output partition, independent of the input size. Some systems use $M \times N$ dedicated single-producer single-consumer (SPSC) channels instead of $N$ multi-producer channels, trading higher memory for lower per-channel contention.

The key advantage over batch partitioning is latency: the first output batch is available as soon as the first input batch is routed, enabling pipelining with downstream operators. The disadvantage is synchronization cost. Each push and pull operation requires coordination---typically via futex, mutex, or atomic compare-and-swap---and with $M$ producers competing on each of $N$ multi-producer single-consumer (MPSC) channels, contention on individual channels grows with thread count. The total synchronization cost per input batch is $O(N)$: one channel operation per output partition.

\subsubsection{Producer-Buffer Variant.}
\label{sec:producer-buffer}
A natural alternative to shared MPSC channels is a \emph{producer-buffer} model, in which each producer maintains $N$ private single-consumer output buffers---one per partition. Consumers poll all $M$ producers' buffers for their partition, yielding $M \times N$ independent SPSC channels. Because each channel has exactly one writer and one reader, synchronization can be entirely lock-free (e.g., a single-producer single-consumer ring buffer~\cite{Baumstark19}), eliminating the multi-producer contention that limits the MPSC channel design. The tradeoff is threefold. First, memory grows to $O(M \times N)$ channel instances, which at 192 producers and 192 consumers means 36{,}864 buffers---each requiring its own allocation and bookkeeping. Second, consumers must visit $M$ separate buffers per batch-group cycle, and the data in those buffers was written by $M$ different producer cores at uncorrelated times, so consumers lose the temporal L3 sharing that the ring-buffer design provides (\autoref{sec:cache-locality}). Third, consumers must implement a polling or notification strategy across $M$ sources, adding per-consumer coordination complexity. The producer-buffer model thus trades lower write-path contention for higher memory, weaker cache locality on the read path, and more complex consumer scheduling. We did not benchmark this variant; a quantitative comparison is an interesting direction for future work.

\subsection{Ring-Buffer Streaming}
\label{sec:design-ring}

\begin{figure}[t]
\centering
\begin{tikzpicture}[
  >=Stealth, scale=0.92, every node/.style={scale=0.92},
  producer/.style={draw, minimum width=1.0cm, minimum height=0.6cm, font=\small},
  consumer/.style={draw, minimum width=1.0cm, minimum height=0.6cm, font=\small},
  group/.style={draw, thick, minimum width=1.8cm, minimum height=0.7cm, rounded corners=3pt},
  slot/.style={draw, minimum width=0.4cm, minimum height=0.45cm, inner sep=0pt},
  lbl/.style={font=\small\itshape, text=gray},
]
  \node[producer] (p1) at (0, 1.0) {$P_1$};
  \node[producer] (p2) at (0, 0.0) {$P_2$};
  \node[producer] (p3) at (0, -1.0) {$P_3$};

  \node[group, fill=yellow!15] (bg) at (3.5, 0) {};
  \node[slot, fill=green!40] at (3.0, 0) {};
  \node[slot, fill=green!40] at (3.5, 0) {};
  \node[slot, fill=white]    at (4.0, 0) {};
  \node[lbl, anchor=north] at (3.5, -0.55) {\texttt{fetch\_add}};
  \node[lbl, anchor=south] at (3.5, 0.55) {current group (filling)};

  \draw[->, thin] (p1.east) -- ++(0.7,0) |- (bg.west);
  \draw[->, thin] (p2.east) -- (bg.west);
  \draw[->, thin] (p3.east) -- ++(0.7,0) |- (bg.west);

  \draw[->, thick, dashed] (3.5, -1.0) -- (3.5, -1.7)
    node[midway, right=3pt, lbl] {publish};

  \node[group, fill=blue!10] (r1) at (3.5, -2.3) {};
  \node[slot, fill=green!40] at (3.0, -2.3) {};
  \node[slot, fill=green!40] at (3.5, -2.3) {};
  \node[slot, fill=green!40] at (4.0, -2.3) {};

  \node[group, fill=blue!10] (r2) at (3.5, -3.2) {};
  \node[slot, fill=green!40] at (3.0, -3.2) {};
  \node[slot, fill=green!40] at (3.5, -3.2) {};
  \node[slot, fill=green!40] at (4.0, -3.2) {};

  \node[group, fill=gray!8] (r3) at (3.5, -4.1) {};
  \node[slot, fill=white] at (3.0, -4.1) {};
  \node[slot, fill=white] at (3.5, -4.1) {};
  \node[slot, fill=white] at (4.0, -4.1) {};

  \draw[decorate, decoration={brace, amplitude=5pt, mirror}, thick]
    (2.3, -1.95) -- (2.3, -4.45)
    node[midway, left=6pt, font=\small, align=right] {ring\\[-2pt]($K$ slots)};

  \node[consumer, fill=blue!15] (c1) at (7.0, -2.3) {$C_1$};
  \node[consumer, fill=red!15]  (c2) at (7.0, -3.2) {$C_2$};

  \draw[->, thick, blue!60] (r1.east) -- (c1.west);
  \draw[->, thick, red!60]  (r2.east) -- (c2.west);

  \node[lbl] at (5.4, -2.0) {read};
  \node[lbl] at (5.4, -2.9) {read};
\end{tikzpicture}
\caption{Ring-buffer streaming. Producers acquire slots in the current batch group via atomic \texttt{fetch\_add}. Full groups are published to the ring buffer ($K$ slots). Each consumer independently tracks its read position and advances through the ring. When consumers are faster than producers, they converge on the same group, enabling temporal cache sharing. Filled slots shown in green; empty in white.}
\Description{Diagram showing three producers filling a batch group, a ring buffer of three groups below, and two consumers each reading from a different ring group.}
\label{fig:ring-design}
\end{figure}
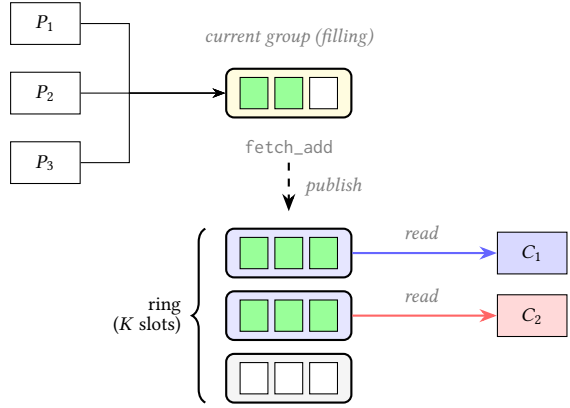

Oxla's ring-buffer design retains the streaming property of channel-based exchange while achieving amortized $O(1)$ synchronization cost per batch and temporal cache locality across consumers. The key idea is to replace per-partition channels with a single shared ring buffer of fixed-size \emph{batch groups}, where producers write concurrently using lock-free atomic counters and consumers independently advance through completed groups---drawing on the sequenced ring-buffer pattern introduced by the LMAX Disruptor~\cite{Thompson11}. \autoref{fig:ring-design} illustrates the architecture. \autoref{fig:pseudocode} summarizes both producer and consumer algorithms in pseudocode.

\subsubsection{Indexed Batches.}
As in channel-based streaming, each producer constructs an indexed batch after the common batch indexing step. The critical difference is how this indexed batch reaches consumers. In channel-based streaming, the producer pushes the indexed batch to $N$ separate channels, incurring a synchronization operation per channel. In the ring-buffer design, the producer places the indexed batch into a single shared structure---a \emph{batch group}---and all consumers read from it there.

\subsubsection{Batch Groups.}
A \emph{batch group} is a fixed-capacity array of $G$ slots, each holding one indexed batch. The capacity $G$ is set proportional to $M$ (the number of producers), so that each producer can contribute approximately one batch before the group fills. Producers acquire a slot by performing a single \texttt{fetch\_add} on an atomic counter---this is the core lock-free operation of the design. If the returned value exceeds $G$, the group is already full and the producer retries with the next group. Otherwise, the producer writes its indexed batch to the acquired slot without any further synchronization. When a producer fills the final slot in a group, it becomes the \emph{publisher} for that group: it pushes the completed group into the ring buffer and installs a fresh group for subsequent writes.

\subsubsection{Data Structure.}

The ring-buffer shuffle consists of a shared queue, per-producer state, and per-consumer state.

\emph{Batch group.} A batch group is a fixed-capacity array of $G$ slots, each holding one indexed batch, plus three atomic counters: \emph{writes\_started}, incremented via \texttt{fetch\_add} to claim a slot; \emph{writes\_completed}, incremented after writing, which triggers the full flag when it reaches $G$; and \emph{consumers\_left}, initialized to $N$ and decremented by each consumer when it finishes reading the group. An atomic boolean full flag signals producers that the group is exhausted.

\emph{Shared state.} The shuffle queue holds a circular ring buffer of $K$ batch group slots (where $K$ is a small constant, typically 1--3), an atomic published counter tracking how many groups have been published, the current \emph{insertion buffer} (the batch group being filled by producers), and a mutex with condition variables for publish, consumer blocking, and producer backpressure. Ring position and occupancy are protected by the mutex.

\emph{Per-producer state.} Each producer holds a private reference to the current insertion buffer, protected by a per-producer mutex. This avoids a shared-pointer bottleneck: when the publisher installs a new insertion buffer, it updates each producer's reference individually, so producers wake and acquire only their own lock---eliminating cross-producer contention. Each producer also holds a pre-allocated replacement batch group, ready to donate when it triggers a publish.

\emph{Per-consumer state.} Each consumer tracks its read position (an absolute index into the sequence of published groups) and a cached copy of the published counter. The cached copy avoids atomic loads on the hot path: the consumer first compares its position against the local cache, performs an atomic load only when caught up, and blocks on a condition variable only when the atomic load also shows no new groups.

\subsubsection{Producer Algorithm.}
\label{sec:producer-algorithm}

When a producer has an indexed batch to submit, it proceeds as follows:

\begin{enumerate}
\item Check the full flag on the batch group referenced by its private pointer. If the group is full, wait on the per-producer condition variable until the publisher installs a new group and updates this producer's reference.
\item Claim a slot via atomic \texttt{fetch\_add} on writes\_started. If the returned index exceeds $G$, the group filled concurrently---return to step~1.
\item Write the indexed batch to the claimed slot. No synchronization is needed for the write itself.
\item Increment writes\_completed. If this was the $G$-th completion:
  \begin{itemize}
  \item Set the full flag.
  \item Acquire the queue mutex. If all $K$ ring slots are occupied, block until a consumer frees a slot (\emph{backpressure}).
  \item Push the full group into the ring, install the pre-allocated replacement as the new insertion buffer, and increment the published counter.
  \item Update all producers' private references to point to the new buffer, and notify waiting consumers and producers.
  \item After releasing the mutex, allocate a fresh replacement group---off the critical path.
  \end{itemize}
\end{enumerate}

The total number of batches in flight is bounded by $K \times G$, independent of the input size.

\subsubsection{Consumer Algorithm.}
\label{sec:consumer-algorithm}

Each consumer independently advances through published batch groups:

\begin{enumerate}
\item Wait for a new batch group. The consumer checks a three-tier progression of increasing cost: first a cached local copy of the published counter (no shared access), then a single atomic load to refresh it, and finally a condition variable wait. This eliminates shared-state access entirely when groups are buffered ahead.
\item Access the batch group at ring position $\textit{read} \bmod K$ and iterate over its indexed batches, extracting the rows designated for this consumer's partition.
\item Decrement consumers\_left. If this consumer is the last reader (count reaches zero): free the ring slot and notify producers if ring occupancy drops below a threshold (e.g., half capacity), allowing multiple slots to accumulate before producers wake. This avoids excessive context switching when consumers are slower than producers.
\end{enumerate}

Different consumers may be reading different batch groups at any given time. When consumers are faster than producers and no new group is available, all waiting consumers block on the same condition variable and are woken simultaneously when the next group is published.

\begin{figure}[t]
\scriptsize
\begin{minipage}[t]{0.48\linewidth}
\textbf{Producer}($\mathit{batch}$)
\hrule\vspace{2pt}
\begin{tabbing}
\hspace{1.2em}\=\hspace{1.2em}\=\kill
1:\> \textbf{loop} \\
2:\>\> \textbf{if} $\mathit{group}$.\textsf{full} \textbf{then wait}(cv$_\mathit{prod}$) \\
3:\>\> $s \gets$ \textsf{fetch\_add}($\mathit{group}$.\textsf{started}, 1) \\
4:\>\> \textbf{if} $s < G$ \textbf{then break} \\
5:\> $\mathit{group}$.\textsf{slots}[$s$] $\gets$ $\mathit{batch}$ \\
6:\> $c \gets$ \textsf{fetch\_add}($\mathit{group}$.\textsf{completed}, 1) \\
7:\> \textbf{if} $c + 1 = G$ \textbf{then} \hfill\emph{// publish} \\
8:\>\> $\mathit{group}$.\textsf{full} $\gets$ \textbf{true} \\
9:\>\> \textsf{lock}($\mu$) \\
10:\>\> \textbf{while} ring is full: \textbf{wait}(cv$_\mathit{bp}$) \\
11:\>\> ring.\textsf{push}($\mathit{group}$) \\
12:\>\> $\mathit{group} \gets$ \textsf{replacement} \\
13:\>\> \textsf{notify}(cv$_\mathit{cons}$, cv$_\mathit{prod}$) \\
14:\>\> \textsf{unlock}($\mu$)
\end{tabbing}
\end{minipage}%
\hfill
\begin{minipage}[t]{0.48\linewidth}
\textbf{Consumer}($\mathit{id}$)
\hrule\vspace{2pt}
\begin{tabbing}
\hspace{1.2em}\=\hspace{1.2em}\=\kill
1:\> \textbf{loop} \\
2:\>\> \textbf{if} $\mathit{pos} <$ cached\_pub \textbf{then break} \\
3:\>\> cached\_pub $\gets$ \textsf{load}(\textsf{published}) \\
4:\>\> \textbf{if} $\mathit{pos} <$ cached\_pub \textbf{then break} \\
5:\>\> \textbf{wait}(cv$_\mathit{cons}$) \\
6:\> $g \gets$ ring[$\mathit{pos} \bmod K$] \\
7:\> \textbf{for each} slot $b$ \textbf{in} $g$ \\
8:\>\> \textsf{extract}($b$, $\mathit{id}$) \\
9:\> $r \gets$ \textsf{fetch\_sub}($g$.\textsf{consumers\_left}, 1) \\
10:\> \textbf{if} $r = 1$ \textbf{then} \hfill\emph{// last reader} \\
11:\>\> ring.\textsf{free}($\mathit{pos} \bmod K$) \\
12:\>\> \textbf{if} occ.\ $\leq K/2$: \textsf{notify}(cv$_\mathit{bp}$) \\
13:\> $\mathit{pos} \gets \mathit{pos} + 1$
\end{tabbing}
\end{minipage}
\caption{Ring-buffer pseudocode. \textbf{Left:} per-batch producer path (lines 1--6) requires one lock-free \textsf{fetch\_add}; the publish path (lines 7--14) acquires a mutex but runs only once per $G$ batches. \textbf{Right:} the consumer is lock-free or local except for the condition variable wait on line~5, which is reached only when caught up to producers.}
\Description{Side-by-side pseudocode for producer and consumer algorithms in the ring-buffer shuffle.}
\label{fig:pseudocode}
\end{figure}

\subsubsection{Synchronization Analysis.}
The producer \emph{hot path}---slot acquisition within a batch group---requires a single atomic \texttt{fetch\_add}, which is $O(1)$ and lock-free. The \emph{cold path}---publishing a full batch group to the ring---requires one mutex acquisition per $G$ batches. Since $G \propto M$, if each of the $M$ producers generates one batch per time unit, the group fills once per time unit regardless of $M$, yielding a constant mutex acquisition rate. By contrast, in channel-based streaming each batch requires a lock per channel, so the total lock acquisition rate is $O(M)$ per time unit---contention grows linearly with producer count. The ring-buffer design eliminates this scaling.

On the consumer side, the fast path is a single atomic read of the published group counter---if a new group is available, the consumer proceeds with no lock. When consumers are faster than producers and no new group is available, the contention on the condition variable is benign: the consumers are idle anyway, and once woken they operate independently on disjoint slices of the same data without further synchronization.

\subsubsection{Practical Considerations.}
\label{sec:practical-considerations}

The asymptotic properties above---amortized $O(1)$ synchronization, $O(M)$ memory---describe the design but do not guarantee good performance on many-core hardware. Two implementation techniques are necessary to realize these properties in practice.

\emph{Pre-allocated replacement groups.}
Publishing a full batch group requires installing a fresh empty group for producers to write into. Allocating this group under the queue mutex---initializing $G$ slots---extends the critical section unnecessarily, increasing the window during which other threads are blocked. Instead, each producer maintains a pre-allocated replacement group. When a producer happens to fill the final slot and triggers a publish, it donates its replacement via a pointer swap (zero-cost under the mutex). After releasing the mutex, it allocates a fresh replacement for next time---off the critical path.

\emph{Selective producer notification.}
When consumers are slower than producers, producers fill the ring and block waiting for space. If a producer is woken as soon as a single slot frees, it fills that slot and immediately blocks again---each group published triggers a round of context switches between producers and consumers for minimal useful work. Instead, producers are notified only when ring occupancy drops to a threshold (e.g., half capacity), allowing multiple slots to accumulate before producers wake. This gives producers enough work to justify the context-switch cost and lets consumers run uninterrupted for longer stretches.

\subsubsection{Memory Analysis.}
The ring buffer holds at most $K \times G = O(M)$ indexed batches in flight. Per-producer private state is $O(1)$: one batch under construction. The total memory footprint is bounded by the ring capacity and proportional to the number of producers, independent of input size. This contrasts with batch partitioning, where memory is $O(|input|)$.

\subsubsection{Cache Locality.}
\label{sec:cache-locality}

The ring-buffer design provides a form of temporal cache locality that neither alternative achieves. Each consumer independently tracks its position in the ring, so different consumers may be processing different batch groups at any given time. However, when consumers are faster than producers---a common scenario when shuffle is the performance bottleneck---consumers that finish a batch group quickly have no new group to advance to and block until the next group is published. At that point, all waiting consumers are notified simultaneously and converge on the same batch group. They each read the same underlying batch data, extracting only their partition's rows via the index. The first consumer to touch a batch pulls its columns into a shared cache level (typically L3); subsequent consumers find the data already warm, paying only L3 hit latency rather than main memory latency. In channel-based streaming, consumers also hold references to the same indexed batches---the data is not copied. However, each consumer pulls from its own channel independently, so consumers process the same batch at different, uncorrelated times. Without the synchronized wake-up that the ring-buffer provides, temporal clustering is weaker and cache sharing is less reliable. In batch partitioning, consumers read data long after producers wrote it (post-barrier), by which time the data has typically been evicted, especially on many-core systems where the aggregate working set exceeds L3 capacity.

\subsubsection{Partition Key Skew.}
\label{sec:skew}
When the partitioning function $h$ distributes rows unevenly (e.g., a small number of hot keys dominate the input), some consumers receive substantially more data than others. In the ring-buffer design, all $N$ consumers must finish reading a batch group before its ring slot can be reclaimed (\texttt{consumers\_left} must reach zero). A slow consumer handling a hot partition therefore gates slot reclamation for the entire group, reducing the effective ring capacity and increasing the likelihood that producers stall on backpressure. This effect is not unique to the ring buffer: in channel-based streaming, backpressure on a hot channel similarly stalls producers that hash to the same partition, and in batch partitioning, the barrier-to-merge latency is determined by the slowest consumer. The ring-buffer design is arguably better positioned than batch partitioning for skewed workloads, because streaming allows fast consumers to proceed immediately rather than waiting at a barrier, and backpressure from a slow consumer affects only the publication rate---not the ability of other consumers to process already-published groups. Nevertheless, extreme skew can degrade ring-buffer throughput, and workloads with known hot keys may benefit from pre-splitting hot partitions or using skew-aware partitioning functions upstream of the shuffle.

\section{Evaluation}
\label{sec:evaluation}

The ring-buffer design relies on atomic counters shared across all producer threads. The efficiency of these atomic operations depends critically on whether the underlying cache line can be transferred between cores through a shared last-level cache or must traverse a slower interconnect between separate cache domains. We therefore select as our primary evaluation platform a dual-socket AWS Graviton4 system (c8g.metal-48xl) with $2{\times}96$ Arm Neoverse V2 cores (192 total), where each socket provides a 36~MB L3 shared across its 96 cores via the Arm CMN-700 mesh. In \autoref{sec:eval-cache-topology}, we contrast these results with a 72-core single-socket Neoverse V2 server (NVIDIA GraceHopper, 117~MB unified LLC) and a chiplet-based AMD EPYC processor to quantify the impact of cache topology on each design.

We evaluate the three shuffle designs using a standalone C++ benchmark\footnote{\url{https://github.com/redpanda-data/ring_shuffle}}. Each experiment uses $M{=}N$ producer-consumer threads, 8192 rows per chunk, and 1000 chunks per producer. All results report the median of 5 runs; variance across runs was low (coefficient of variation below 5\% for all configurations). Producers and consumers are pinned to the same physical cores to simulate a fixed core budget, as is typical in database engines. Note that this approach simulates limited compute but not limited memory bandwidth---lower-core-count systems typically also have proportionally lower memory bandwidth, which may shift the crossover point between synchronization-bound and bandwidth-bound regimes. All three implementations were developed by the authors with equal optimization effort; the benchmark code is available in the artifact repository. The Channel baseline uses one bounded MPSC queue per output partition ($N$ total), each backed by a \texttt{std::vector} under a \texttt{std::mutex} with separate condition variables for not-full and not-empty; capacity is fixed at $M$ batches per partition. The Batch baseline gives each producer $N$ thread-local \texttt{std::vector} buckets that hold indexed-batch pointers; producers append without synchronization, and after all producers join, consumers iterate across all $M$ producer buckets to extract their partition. All three designs shuffle indexed-batch pointers rather than copying row payloads, so throughput differences reflect coordination cost rather than data movement.

\subsection{Scalability}
\label{sec:eval-scalability}

\begin{figure}[t]
\centering
\begin{tikzpicture}
\begin{axis}[
  width=\linewidth,
  height=5.5cm,
  xlabel={Thread count ($M{=}N$)},
  ylabel={Throughput (GB/s)},
  xmode=log,
  log basis x=2,
  xtick={1,2,4,8,16,32,64,128,192},
  xticklabels={1,2,4,8,16,32,64,128,192},
  ymin=0, ymax=14,
  legend style={at={(0.03,0.97)}, anchor=north west, font=\small},
  grid=major,
  grid style={gray!30},
  every axis plot/.append style={thick, mark size=2pt},
]
\addplot[color=blue!70, mark=square*] coordinates {
  (1,0.34) (2,0.64) (4,1.17) (8,1.97) (16,2.82) (32,3.33) (64,3.49) (128,3.67) (192,2.94)
};
\addplot[color=red!70, mark=triangle*] coordinates {
  (1,0.41) (2,0.74) (4,1.43) (8,2.65) (16,4.73) (32,5.80) (64,7.42) (128,5.85) (192,6.44)
};
\addplot[color=black!80, mark=*, mark options={fill=black}] coordinates {
  (1,0.42) (2,0.81) (4,1.57) (8,2.97) (16,5.27) (32,5.86) (64,10.18) (128,10.70) (192,11.62)
};
\legend{Batch, Channel, Ring}
\end{axis}
\end{tikzpicture}
\caption{Throughput scaling with thread count on a 192-core dual-socket Graviton4 (Neoverse V2). Row size is 8~bytes (uniform). Beyond 96 cores (socket boundary), channel throughput drops while ring-buffer streaming continues to scale; batch saturates early.}
\Description{Line chart showing throughput vs thread count from 1 to 192 for three shuffle algorithms. Ring buffer scales to 11.62 GB/s at 192 cores. Channel peaks at 7.42 GB/s at 64 cores then drops at 128 before recovering to 6.44 at 192. Batch plateaus around 3.5 GB/s.}
\label{fig:scalability}
\end{figure}
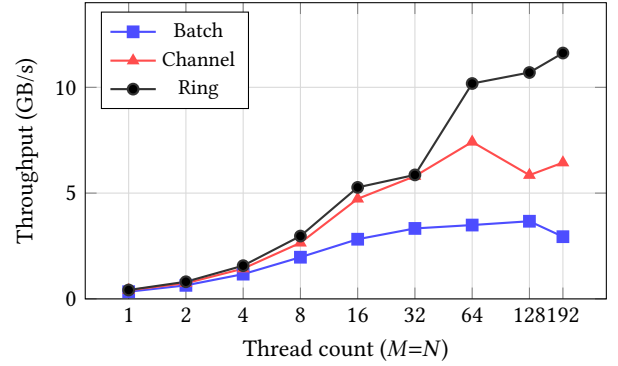

\autoref{fig:scalability} shows throughput as thread count increases from 1 to 192 on the dual-socket Graviton4, with 8-byte rows and uniform row sizes. At low core counts ($\leq 8$), all three approaches perform similarly---contention on shared synchronization state is minimal, so the ring-buffer's mechanisms for avoiding it provide no benefit. Starting at 16 cores, contention becomes the dominant cost and the ring-buffer design ($K{=}1$) begins to pull ahead. At 64 cores (within a single socket), ring-buffer streaming achieves 10.18~GB/s, compared to 7.42~GB/s for channel-based streaming ($+37\%$) and 3.49~GB/s for batch partitioning ($+192\%$). Beyond 96 cores, threads span both sockets; channel throughput drops from 7.42~GB/s at 64 cores to 5.85~GB/s at 128 as cross-socket synchronization costs increase, while ring-buffer throughput continues to rise. At 192 cores, ring-buffer streaming achieves 11.62~GB/s, compared to 6.44~GB/s for channel-based streaming ($+80\%$) and 2.94~GB/s for batch partitioning ($+295\%$).

\subsection{Row Size and Distribution Sensitivity}
\label{sec:eval-rowsize}

\autoref{tab:rowsize} compares the three approaches at 192 cores across batch sizes from 64~KB to 2048~KB (8192 rows with row sizes from 8 to 256~bytes), under both uniform and normal ($\mu{=}\text{row\_size}$, $\sigma{=}\mu/4$) row-size distributions. For each configuration, we report ring-buffer throughput ($K{=}1$) alongside the percentage improvement over batch and channel. The highest throughput per configuration is shown in bold. At 1024~KB and 2048~KB, batch partitioning exceeds the 128~GB memory limit and cannot run---a direct consequence of its $O(|input|)$ memory footprint at 192 producers.

The ring-buffer advantage is largest at small batch sizes: at 64~KB with uniform rows, ring-buffer streaming achieves $+253\%$ over batch and $+106\%$ over channel. Small batches are processed quickly, causing producers and consumers to hit synchronization points more frequently per unit time---maximizing contention on locks and condition variables. As batch size increases, each chunk takes longer to process, reducing the synchronization rate, and the workload shifts from synchronization-bound to memory-bandwidth-bound. At 2048~KB, ring-buffer leads channel by $+67\%$---the gap narrows but remains substantial at 192 cores.

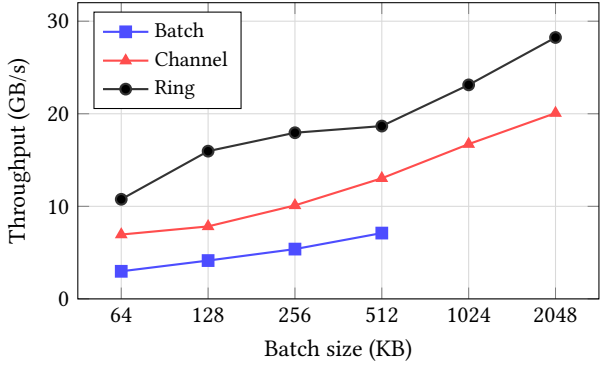
\begin{figure}[t]
\centering
\begin{tikzpicture}
\begin{axis}[
  width=\linewidth,
  height=5.5cm,
  xlabel={Batch size (KB)},
  ylabel={Throughput (GB/s)},
  xmode=log,
  log basis x=2,
  xtick={64,128,256,512,1024,2048},
  xticklabels={64,128,256,512,1024,2048},
  ymin=0, ymax=32,
  legend style={at={(0.03,0.97)}, anchor=north west, font=\small, cells={anchor=west}},
  legend columns=1,
  grid=major,
  grid style={gray!30},
  every axis plot/.append style={thick, mark size=2pt},
]
\addplot[color=blue!70, mark=square*] coordinates {
  (64,2.98) (128,4.14) (256,5.38) (512,7.11)
};
\addplot[color=red!70, mark=triangle*] coordinates {
  (64,6.95) (128,7.83) (256,10.10) (512,13.03) (1024,16.72) (2048,20.07)
};
\addplot[color=black!80, mark=*, mark options={fill=black}] coordinates {
  (64,10.76) (128,15.96) (256,17.95) (512,18.67) (1024,23.12) (2048,28.24)
};
\legend{Batch, Channel, Ring}
\end{axis}
\end{tikzpicture}
\caption{Throughput vs.\ batch size at 192 cores (Graviton4) with ring-buffer $K{=}1$ and normally distributed row sizes. Batch partitioning exceeds memory at $\geq$1024~KB. The ring-buffer advantage is largest at small batch sizes where synchronization dominates.}
\Description{Line chart showing throughput vs batch size for three shuffle algorithms at 192 cores with normally distributed row sizes. Ring buffer leads across all batch sizes, reaching 28.24 GB/s at 2048 KB. Batch line stops at 512 KB due to memory limits.}
\label{fig:batchsize}
\end{figure}

\autoref{fig:batchsize} visualizes this trend under normal row-size distribution, which better reflects real-world workloads where row sizes vary. The ring-buffer curve remains above the other two across all batch sizes.

The normal distribution shows slightly lower absolute throughput. Variable row sizes cause uneven batch sizes across partitions: some consumers receive more data than others, leading to stragglers. In streaming approaches, faster consumers back-pressure their producers while waiting for slower consumers to catch up. In batch partitioning, the barrier amplifies the imbalance---the slowest consumer determines the completion time of the merge phase. Despite this, the relative advantage of ring-buffer streaming is preserved across both distributions.

\begin{table}[t]
\caption{Throughput (GB/s) at 192 cores (Graviton4). Ring-buffer uses $K{=}1$. Batch size is rows $\times$ row size (8192 rows). Percentage gain of ring-buffer over each baseline shown in parentheses. Best throughput per configuration bolded. ``---'' indicates out-of-memory ($>$128~GB).}
\label{tab:rowsize}
\small
\setlength{\tabcolsep}{3.5pt}
\begin{tabular}{clrrr}
\toprule
\textbf{Batch (KB)} & \textbf{Dist.} & \textbf{Batch} & \textbf{Channel} & \textbf{Ring $K{=}1$} \\
\midrule
64    & uniform &  3.29 \scriptsize{(+253\%)} &  5.64 \scriptsize{(+106\%)} & \textbf{11.61} \\
128   & uniform &  3.97 \scriptsize{(+335\%)} &  9.17 \scriptsize{(+88\%)}  & \textbf{17.26} \\
256   & uniform &  5.26 \scriptsize{(+300\%)} & 11.41 \scriptsize{(+84\%)}  & \textbf{21.04} \\
512   & uniform &  7.30 \scriptsize{(+301\%)} & 15.26 \scriptsize{(+92\%)}  & \textbf{29.31} \\
1024  & uniform & ---                         & 19.05 \scriptsize{(+85\%)}  & \textbf{35.17} \\
2048  & uniform & ---                         & 23.85 \scriptsize{(+67\%)}  & \textbf{39.82} \\
\midrule
64    & normal &  2.98 \scriptsize{(+261\%)} &  6.95 \scriptsize{(+55\%)}  & \textbf{10.76} \\
128   & normal &  4.14 \scriptsize{(+286\%)} &  7.83 \scriptsize{(+104\%)} & \textbf{15.96} \\
256   & normal &  5.38 \scriptsize{(+234\%)} & 10.10 \scriptsize{(+78\%)}  & \textbf{17.95} \\
512   & normal &  7.11 \scriptsize{(+163\%)} & 13.03 \scriptsize{(+43\%)}  & \textbf{18.67} \\
1024  & normal & ---                         & 16.72 \scriptsize{(+38\%)}  & \textbf{23.12} \\
2048  & normal & ---                         & 20.07 \scriptsize{(+41\%)}  & \textbf{28.24} \\
\bottomrule
\end{tabular}
\end{table}

\subsection{Impact of Cache Topology}
\label{sec:eval-cache-topology}

The results in \autoref{sec:eval-scalability} and \autoref{sec:eval-rowsize} were measured on the dual-socket Graviton4 system described in \autoref{sec:evaluation}. To evaluate the impact of cache topology, we compare with two additional platforms. First, a 72-core single-socket Arm Neoverse V2 server (NVIDIA GraceHopper) with a unified 117~MB LLC shared by all cores---a single coherence domain. Second, an AMD EPYC 7R13 server with 96 physical cores (192 vCPUs via SMT), which uses a chiplet architecture: 12 Core Complex Dies (CCDs), each containing 8 cores with a private 32~MB L3 cache. There is no shared cache across CCDs---cross-CCD coherence is maintained via a directory-based protocol over the Infinity Fabric interconnect, with measured latency increasing significantly across CCD boundaries~\cite{Fogli25}.

\autoref{tab:topology} compares ring-buffer and channel-based streaming at full core count on all three architectures. On the 72-core Grace system, ring-buffer streaming outperforms channel streaming at every batch size, with gains up to $+44\%$. On the 192-core Graviton4, the ring-buffer advantage is substantially larger---up to $+106\%$ at 64~KB---despite the dual-socket NUMA topology, because the atomic counter bounces between only 2 sockets rather than 12 CCDs. The higher core count amplifies channel contention, making the ring buffer's amortized $O(1)$ synchronization rate more valuable. Additionally, the Graviton4's higher aggregate memory bandwidth (dual-socket DDR5) raises the bandwidth ceiling, keeping the workload in the synchronization-bound regime across all tested batch sizes. On Grace and EPYC, memory bandwidth saturates at smaller batch sizes, compressing the gap between approaches---on EPYC this effect is particularly pronounced, as 192 vCPUs sharing the available bandwidth achieve lower peak throughput (18.42~GB/s at 2048~KB) than Grace with only 72 cores (21.79~GB/s). Notably, peak shuffle throughput at 2048~KB is roughly proportional to each platform's theoretical memory bandwidth---Grace (LPDDR5X, 546~GB/s), Graviton4 (dual-socket DDR5, ${\sim}$1,075~GB/s), and EPYC (dual-socket DDR4, ${\sim}$410~GB/s)---and Grace achieves the highest per-core shuffle throughput (0.30~GB/s versus 0.21 for Graviton4 and 0.10 for EPYC), consistent with its 7.6~GB/s per-core theoretical bandwidth. On the EPYC, the picture is mixed: channel streaming matches or exceeds the ring buffer at 256~KB, 512~KB, and 2048~KB batches, while the ring buffer regains the lead at 1024~KB---likely due to interaction between batch size and CCD-local cache capacity.

We hypothesize that the performance difference stems from how atomic operations are resolved on each architecture. On the Grace system, atomic operations on shared counters (slot acquisition, completion tracking) resolve within the unified LLC at low latency. On the EPYC, no shared cache exists across CCDs. Ownership of the cache line holding the ring buffer's atomic slot counter must transfer between CCDs via the Infinity Fabric, with each transfer incurring significantly higher latency than an intra-CCD operation~\cite{Fogli25}. Since the ring buffer funnels all slot acquisition through a single atomic counter, this cache line bounces across 12 CCDs on every group fill, serializing producers at interconnect speed---a pattern analogous to the timestamp counter scalability collapse observed by Yu et al.~\cite{Yu14}. The EPYC's mixed results thus stem from two compounding factors: cross-CCD atomic bouncing penalizes the ring buffer's single-counter design, while early bandwidth saturation limits the throughput ceiling for all approaches, leaving little room for the ring buffer's lower synchronization overhead to translate into gains.

These results suggest that the ring-buffer design is best suited for architectures with \emph{few coherence domains}---whether a single unified LLC (Grace) or a small number of sockets with per-socket shared caches (Graviton4). The critical factor is not a single unified cache per se, but the number of domains across which the atomic counter must bounce: 1 domain (Grace) and 2 domains (Graviton4) both strongly favor the ring buffer, while 12 domains (EPYC CCDs) erode the advantage. On multi-CCD architectures such as AMD EPYC and multi-tile Intel Xeon, channel-based streaming remains competitive, particularly at larger batch sizes where the workload is memory-bandwidth-bound. A NUMA-aware variant of the ring-buffer design---using per-CCD batch groups with local atomic counters---could potentially recover the advantage on chiplet architectures but is beyond the scope of this work.

\begin{table}[t]
\caption{Ring vs.\ channel throughput (GB/s) across cache topologies. Grace: 72 Neoverse V2 cores, unified 117~MB LLC. Graviton4: $2{\times}96$ Neoverse V2 cores, $2{\times}36$~MB L3 (dual socket). EPYC Milan: 192 vCPUs (96 cores + SMT), 12 CCDs with 32~MB L3 each. All use ring $K{=}1$. Best per configuration bolded.}
\label{tab:topology}
\footnotesize
\setlength{\tabcolsep}{2pt}
\newcommand{\tW}[1]{\makebox[3.2em][r]{#1}}
\begin{tabular}{crrrrrr}
\toprule
 & \multicolumn{2}{c}{\textbf{Grace (72c)}} & \multicolumn{2}{c}{\textbf{Graviton4 (192c)}} & \multicolumn{2}{c}{\textbf{EPYC (192v)}} \\
\cmidrule(lr){2-3} \cmidrule(lr){4-5} \cmidrule(lr){6-7}
\textbf{Batch} & \textbf{Chan.} & \textbf{Ring} & \textbf{Chan.} & \textbf{Ring} & \textbf{Chan.} & \textbf{Ring} \\
\midrule
64\,KB   & \tW{7.87}  & \tW{\textbf{11.30}} & \tW{5.64}  & \tW{\textbf{11.61}} & \tW{3.45}           & \tW{\textbf{3.94}} \\
128\,KB  & \tW{12.33} & \tW{\textbf{15.37}} & \tW{9.17}  & \tW{\textbf{17.26}} & \tW{5.35}           & \tW{\textbf{5.65}} \\
256\,KB  & \tW{15.29} & \tW{\textbf{18.35}} & \tW{11.41} & \tW{\textbf{21.04}} & \tW{\textbf{8.43}}  & \tW{8.26} \\
512\,KB  & \tW{17.24} & \tW{\textbf{18.86}} & \tW{15.26} & \tW{\textbf{29.31}} & \tW{\textbf{12.54}} & \tW{11.85} \\
1024\,KB & \tW{17.73} & \tW{\textbf{19.98}} & \tW{19.05} & \tW{\textbf{35.17}} & \tW{15.87}          & \tW{\textbf{18.01}} \\
2048\,KB & \tW{19.45} & \tW{\textbf{21.79}} & \tW{23.85} & \tW{\textbf{39.82}} & \tW{\textbf{18.42}} & \tW{17.77} \\
\bottomrule
\end{tabular}
\end{table}

\subsection{Effect of Ring Capacity K}
\label{sec:eval-ring-k}

The experiments above all use $K{=}1$, the smallest possible ring buffer. \autoref{sec:design-ring} noted that $K$ is typically 1--3; we now measure how $K$ affects throughput and show that the optimal value depends on cache topology.

\begin{figure}[t]
\centering
\begin{tikzpicture}
\begin{axis}[
  width=\linewidth, height=3.6cm,
  title={Grace (72c)}, title style={font=\small, yshift=-4pt},
  xlabel={}, ylabel={GB/s}, ylabel style={yshift=-4pt},
  xmode=log, log basis x=2,
  xtick={64,128,256,512,1024,2048},
  xticklabels={64,128,256,512,1024,2048},
  ymin=10, ymax=24,
  legend style={at={(0.03,0.97)}, anchor=north west, font=\tiny, draw=none, fill=white, fill opacity=0.6, text opacity=1, cells={anchor=west}},
  legend columns=2,
  grid=major, grid style={gray!30},
  every axis plot/.append style={thick, mark size=1.5pt},
]
\addplot[color=black,     mark=*]         coordinates {(64,11.30) (128,15.37) (256,18.35) (512,18.86) (1024,19.98) (2048,21.79)};
\addplot[color=blue!70,   mark=square*]   coordinates {(64,13.17) (128,16.80) (256,17.96) (512,18.35) (1024,18.89) (2048,20.61)};
\addplot[color=red!70,    mark=triangle*] coordinates {(64,13.19) (128,16.53) (256,16.80) (512,17.22) (1024,18.50) (2048,20.25)};
\addplot[color=orange!85, mark=diamond*]  coordinates {(64,12.95) (128,15.50) (256,13.88) (512,14.83) (1024,16.99) (2048,20.12)};
\legend{$K{=}1$, $K{=}2$, $K{=}3$, $K{=}4$}
\end{axis}
\end{tikzpicture}

\vspace{-2pt}
\begin{tikzpicture}
\begin{axis}[
  width=\linewidth, height=3.6cm,
  title={Graviton4 (192c)}, title style={font=\small, yshift=-4pt},
  xlabel={}, ylabel={GB/s}, ylabel style={yshift=-4pt},
  xmode=log, log basis x=2,
  xtick={64,128,256,512,1024,2048},
  xticklabels={64,128,256,512,1024,2048},
  ymin=8, ymax=44,
  grid=major, grid style={gray!30},
  every axis plot/.append style={thick, mark size=1.5pt},
]
\addplot[color=black,    mark=*]       coordinates {(64,11.61) (128,17.26) (256,21.04) (512,29.31) (1024,35.17) (2048,39.82)};
\addplot[color=blue!70,  mark=square*] coordinates {(64,10.10) (128,14.73) (256,14.27) (512,19.83) (1024,26.62) (2048,32.74)};
\end{axis}
\end{tikzpicture}

\vspace{-2pt}
\begin{tikzpicture}
\begin{axis}[
  width=\linewidth, height=3.6cm,
  title={EPYC (192v)}, title style={font=\small, yshift=-4pt},
  xlabel={Batch size (KB)}, ylabel={GB/s}, ylabel style={yshift=-4pt},
  xmode=log, log basis x=2,
  xtick={64,128,256,512,1024,2048},
  xticklabels={64,128,256,512,1024,2048},
  ymin=2, ymax=20,
  grid=major, grid style={gray!30},
  every axis plot/.append style={thick, mark size=1.5pt},
]
\addplot[color=black,     mark=*]         coordinates {(64,3.94) (128,5.65) (256,8.26) (512,11.85) (1024,18.01) (2048,17.77)};
\addplot[color=blue!70,   mark=square*]   coordinates {(64,3.55) (128,5.16) (256,7.31) (512,13.54) (1024,16.45) (2048,16.41)};
\addplot[color=red!70,    mark=triangle*] coordinates {(64,3.60) (128,5.16) (256,7.22) (512,12.60) (1024,15.54) (2048,15.91)};
\addplot[color=orange!85, mark=diamond*]  coordinates {(64,3.52) (128,5.01) (256,7.21) (512,12.23) (1024,15.37) (2048,16.27)};
\end{axis}
\end{tikzpicture}
\caption{Ring-buffer throughput vs.\ batch size at full core count, sweeping $K \in \{1,2,3,4\}$ (Graviton4 measured for $K\in\{1,2\}$ only; uniform row sizes, flat distribution). On Grace, $K{=}2$ and $K{=}3$ win at small batch sizes where stall frequency is high; on Graviton4 and EPYC, additional in-flight groups generate cross-domain cache-line traffic that outweighs the stall-reduction benefit, so $K{=}1$ dominates almost everywhere.}
\Description{Three stacked line charts showing ring-buffer throughput vs.\ batch size for K=1, K=2, K=3, K=4 on Grace (top), Graviton4 (middle, K=1 and K=2 only), and EPYC (bottom).}
\label{fig:ring-k}
\end{figure}
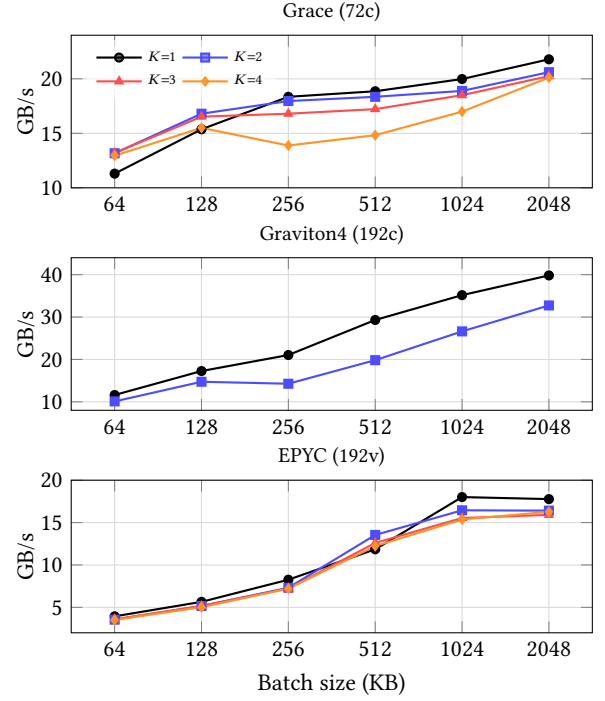

\autoref{fig:ring-k} sweeps $K \in \{1,2,3,4\}$ across batch sizes at full core count on each platform (Graviton4 was measured for $K \in \{1,2\}$ only). $K$ controls the slack between producers and consumers: with $K{=}1$ the ring holds at most one already-published group, so producers stall when the ring is full and they need to publish a fresh group, and consumers stall when no new group has yet been published. The frequency of both stalls scales inversely with per-group time---when each group fills or drains in microseconds, even brief rate variance causes frequent stalls; when each group takes milliseconds, stalls are rare regardless of $K$. Larger $K$ provides additional in-flight groups that absorb this variance, but at the cost of a larger working set across in-flight groups and weaker consumer convergence on shared groups (\autoref{sec:cache-locality}).

\emph{Grace.} At small batch sizes ($\leq 128$\,KB), $K{=}2$ and $K{=}3$ outperform $K{=}1$ by 9--17\%. Two factors align in $K{>}1$'s favor: the aggregate working set across multiple in-flight groups still fits in the 117\,MB L3 (no cache penalty), and per-group times are short enough that stall frequency is high, so the additional buffering pays for itself. Beyond 256\,KB the picture flips: the working set under $K{>}1$ begins to exceed L3, consumers drift onto different groups (breaking the convergence described in \autoref{sec:cache-locality}), and stall frequency drops below the threshold where buffering matters. $K{=}1$ retakes the lead by 2--6\%. $K{=}4$ falls off most steeply at medium batch sizes (256--512\,KB) where the four-group working set is first to exceed L3.

\emph{Graviton4.} $K{=}1$ dominates $K{=}2$ at every batch size, with the gap peaking at 256--512\,KB (32\%) and remaining substantial (18\%) at 2048\,KB. The dual-socket NUMA topology turns every additional in-flight group into cross-socket cache-line traffic on the shared atomic counters and ring bookkeeping; this fixed coordination cost outweighs the stall-reduction benefit observed on Grace, even at small batch sizes.

\emph{EPYC.} $K{=}1$ wins at most batch sizes; the only exception is 512\,KB, where $K{=}2$ leads by 14\% before $K{=}1$ resumes its lead at larger sizes. The chiplet topology's high cross-CCD coherence latency penalizes additional in-flight state, mirroring the Graviton4 pattern.

The rule that emerges: default to $K{=}1$, and tune toward $K{=}2$--$3$ only on architectures with a single coherence domain at small batch sizes, where the high stall frequency makes the additional buffering pay for itself. Production Oxla deployments on Grace-class hardware use $K{=}2$. On chiplet or multi-socket hardware where atomic-counter traffic crosses cache domains, $K{=}1$ is consistently the safe default. Across all three platforms $K{=}4$ is dominated by smaller $K$ values, validating the ``typically 1--3'' design guideline of \autoref{sec:design-ring}.

\subsection{End-to-End Query Impact}
\label{sec:e2e}
\label{sec:eval-tpch}

The microbenchmarks above isolate the shuffle primitive. To measure how that translates to real queries, we compared the ring-buffer build (currently in production at Oxla) against a channel-based build that wires Oxla's hash-aggregation pipeline through one MPSC channel per output partition, with rows hash-routed and row-copied into per-bucket batches before being pushed onto each channel. The channel-based shuffle is not a hypothetical baseline: it was Oxla's production shuffle before the ring-buffer design replaced it. Both builds run on the c8g.metal-48xl Graviton4 from \autoref{sec:evaluation}; ring capacity is $K{=}2$ unless noted. We report median wall-clock from warmed-up runs; coefficient of variation is below 3\% across all measurements. We evaluate two standard analytic suites at the scales they are typically run: TPC-H Q1--Q22 at SF=100 (Q21 omitted---it requires \texttt{EXISTS} subqueries that Oxla does not support) and the full 43-query ClickBench suite at the standard 99.9\,M-row scale.

\autoref{tab:tpch-q18} reports the suite-level results alongside the microbenchmark from \autoref{sec:eval-rowsize}. On TPC-H---21 queries spanning hash joins with multi-table aggregations, sub-queries, complex predicates, and string aggregations---the ring-buffer wins in aggregate (86.6\,s vs 92.9\,s, 1.07$\times$). On ClickBench---43 variations on \emph{filter + GROUP BY + aggregate}---the channel build wins the suite total (9.0\,s vs 13.5\,s, 1.49$\times$), driven by four queries on the \texttt{COUNT(DISTINCT)} and wide-aggregate end.

\begin{table}[t]
\caption{Ring-buffer vs channel-based shuffle on Graviton4 (192 cores). Ring uses $K{=}2$. Microbenchmark row from \autoref{sec:eval-rowsize}; suite rows are warmed-up sums-of-medians (TPC-H 21 queries, ClickBench 43 queries).}
\label{tab:tpch-q18}
\small
\setlength{\tabcolsep}{4pt}
\begin{tabular}{lrrr}
\toprule
\textbf{Workload} & \textbf{Ring} & \textbf{Channel} & \textbf{Ring vs Ch.} \\
\midrule
Microbench (8\,B rows, 64\,KB) & 11.6\,GB/s & 5.6\,GB/s & Ring \textbf{1.6$\times$} \\
TPC-H 21 queries (sum)                                        &  86.6\,s   & 92.9\,s   & Ring \textbf{1.07$\times$} \\
ClickBench 43 queries (sum)                                   &  13.5\,s   &  9.0\,s   & Ch.\ \textbf{1.49$\times$} \\
\bottomrule
\end{tabular}
\end{table}

The two suite-level numbers measure different things. TPC-H spans diverse query shapes---hash joins, sub-queries, multi-table aggregations, string aggregations, predicates of widely varying selectivity---across 21 queries; ClickBench densely samples a narrow band of \emph{filter + GROUP BY + aggregate} variants across 43 queries. The aggregate ratio for TPC-H therefore measures how the two designs compare across diverse query shapes; the aggregate ratio for ClickBench measures how they compare under one dominant shape repeated many ways.

\begin{figure}[t]
\centering
\begin{tikzpicture}
\begin{axis}[
  name=tpchax,
  width=\linewidth, height=4.8cm,
  xmin=0.5, xmax=21.5, ymin=-0.6, ymax=0.7,
  xlabel={}, ylabel={},
  title={TPC-H (21 queries)}, title style={font=\small, yshift=-3pt},
  xtick=\empty, axis x line=middle,
  grid=major, grid style={gray!25},
]
\addplot+[ybar, bar width=4pt, fill=red!60!black, draw=red!60!black] coordinates {
(1,-0.234) (2,-0.224) (3,-0.124) (4,-0.041) (5,-0.038) (6,-0.024) (7,-0.002)};
\addplot+[ybar, bar width=4pt, fill=blue!50!black, draw=blue!50!black] coordinates {
(8,0.032)  (9,0.067)  (10,0.070) (11,0.077) (12,0.086) (13,0.093) (14,0.144)
(15,0.161) (16,0.169) (17,0.178) (18,0.184) (19,0.307) (20,0.468) (21,0.552)};
\end{axis}
\begin{axis}[
  at={(tpchax.below south west)}, anchor=above north west, yshift=-12pt,
  width=\linewidth, height=4.8cm,
  xmin=0.5, xmax=43.5, ymin=-2.6, ymax=1.3,
  xlabel={}, ylabel={},
  title={ClickBench (43 queries)}, title style={font=\small, yshift=-3pt},
  xtick=\empty, axis x line=middle,
  grid=major, grid style={gray!25},
]
\addplot+[ybar, bar width=3pt, fill=red!60!black, draw=red!60!black] coordinates {
(1,-2.30) (2,-2.07) (3,-1.46) (4,-1.16) (5,-1.08) (6,-0.95) (7,-0.87) (8,-0.78)
(9,-0.73) (10,-0.71) (11,-0.69) (12,-0.58) (13,-0.50) (14,-0.25) (15,-0.24) (16,-0.23)
(17,-0.18) (18,-0.18) (19,-0.16) (20,-0.14) (21,-0.13) (22,-0.12) (23,-0.10) (24,-0.04)
(25,-0.03) (26,-0.01)};
\addplot+[ybar, bar width=3pt, fill=blue!50!black, draw=blue!50!black] coordinates {
(27,0.00) (28,0.04) (29,0.04) (30,0.05) (31,0.06) (32,0.07) (33,0.07) (34,0.11)
(35,0.14) (36,0.17) (37,0.24) (38,0.24) (39,0.43) (40,0.59) (41,0.63) (42,0.72) (43,0.84)};
\end{axis}
\end{tikzpicture}
\caption{Per-query $\log_2(t_{\text{channel}}/t_{\text{ring}})$ on Graviton4 (192 cores), sorted. Positive bars: ring-buffer faster. TPC-H bars stay within $\pm 0.55$ on the log axis (max 1.47$\times$ either way) and the bulk are positive. ClickBench is bimodal: a long tail of small ring-buffer wins and ties on the right, plus four queries on the left where the ring-buffer loses by 2--5$\times$.}
\Description{Two stacked bar charts. Top panel (TPC-H): 21 queries sorted by log2 ratio, bulk positive (ring wins), ranging from Q2 at -0.23 to Q22 at +0.55. Bottom panel (ClickBench): 43 queries, bimodal---Q33 at -2.30 and Q14 at -2.07 on the left (channel wins by 4--5x), Q15 at +0.84 on the right (ring wins by 1.8x), many bars near zero in the middle.}
\label{fig:cb-ratio}
\end{figure}
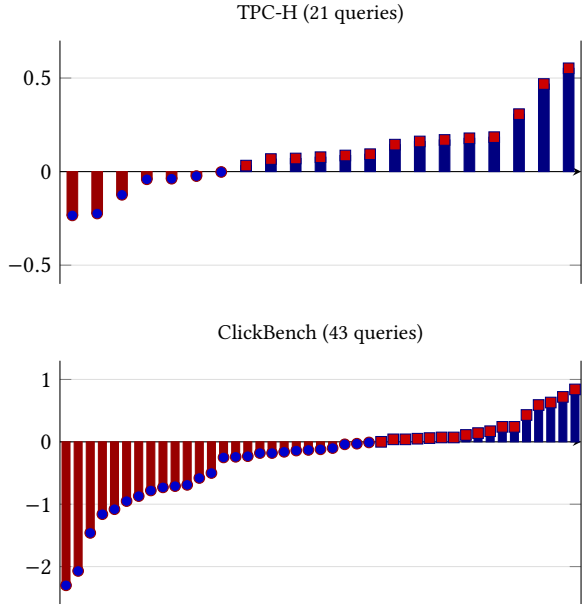

\autoref{fig:cb-ratio} shows the per-query distribution behind those numbers. Each panel plots $\log_2(t_{\text{channel}} / t_{\text{ring}})$ for every query in the suite, sorted: positive bars are queries where the ring-buffer is faster, negative bars where the channel build is faster. The two distributions look very different. TPC-H clusters within roughly $\pm 0.55$ on the log axis (i.e.\ no query is more than 1.47$\times$ apart between the two builds), and the bulk of the queries are positive. ClickBench is bimodal: a long tail of small ring-buffer wins and ties on the right, plus a handful of queries on the left where the ring-buffer loses by a much larger margin than anywhere on the TPC-H panel.

\paragraph{Where the ring-buffer wins, and where it loses.} The ring-buffer's wins concentrate where a query exposes the shuffle as a throughput-bound stage with simple per-group state: hash joins on integer keys with a small set of post-shuffle aggregates (TPC-H Q22, Q8, Q17, Q5, Q9, Q12, Q19), or \texttt{GROUP BY} on a moderate-cardinality string column with a counter aggregate (ClickBench Q15, Q2, Q31, Q13, Q32). The wins are bounded---typically 1.1--1.5$\times$, and only Q22 on TPC-H reaches 1.47$\times$. The ring-buffer's losses concentrate in two narrow query shapes that put heavy per-row work on the consumer side: \texttt{COUNT(DISTINCT)} per group (ClickBench Q14: 4.2$\times$ slower; TPC-H Q13: 1.17$\times$ slower) and wide per-row arithmetic over many simultaneous aggregates (ClickBench Q33: 4.9$\times$ slower for its 90 \texttt{SUM} expressions). When the shuffle's consumer maintains a hash set per group, or evaluates dozens of arithmetic expressions per row before aggregating, the ring-buffer's shared-batch + indexed-read access pattern is unfavorable for the consumer's working set, while the channel build delivers each consumer's rows already grouped contiguously.

The ring-buffer's advantage is therefore distributed as many small wins concentrated in queries that expose its design regime; its disadvantage is concentrated in a few specific query shapes that defeat that regime. Whether a workload nets out positive or negative for the ring-buffer depends on how many queries of each shape it contains. TPC-H, dominated by hash joins on integer keys with simple aggregates, contains many ring-favorable queries and few unfavorable ones; ClickBench contains both, plus a handful of \texttt{COUNT(DISTINCT)} and wide-aggregate queries large enough in absolute time to swing the suite total.

\paragraph{$K$ does not flip the ranking.} We sweep ring capacity $K \in \{1,2\}$ on the ring-buffer build. Across both suites the two settings are within 3\% of each other (TPC-H sums: K=1 83.9\,s, K=2 86.6\,s; ClickBench sums: K=1 13.8\,s, K=2 13.5\,s). The per-query pattern in \autoref{fig:cb-ratio} is structural---a property of the design comparison, not of the parameter setting. Detailed K-vs-batch-size data is in \autoref{sec:eval-ring-k}.

\paragraph{Synthesis.} The TPC-H result is the broad-applicability evidence: across 21 queries that exercise hash joins, sub-queries, multi-table aggregations, string aggregations, and filters of widely varying selectivity, the ring-buffer wins in aggregate, and the wins are distributed across diverse query shapes rather than concentrated in any one. The microbenchmark in \autoref{sec:eval-rowsize}---CRC-only consumers, continuous full-rate production, row size swept from 8 to 256 bytes---describes a regime (light per-row consumer work, no per-group state, throughput-bound shuffle) that recurs inside real queries as the inner loop of a hash join with simple aggregates, or a \texttt{GROUP BY} on a moderate-width key with a counter aggregate. Wherever a query plan exposes that regime to the shuffle, the microbenchmark's prediction lands directly: 1.1--1.5$\times$. The ring-buffer's small but consistent wins across the bulk of TPC-H, and its larger wins on the right tail of ClickBench, both come from this mechanism. The ClickBench suite total swings the other way because ClickBench's narrower shape distribution oversamples a handful of queries that put heavy per-row work on the consumer---\texttt{COUNT(DISTINCT)} per group, wide-row arithmetic over many simultaneous aggregates---where the channel design's per-bucket contiguous delivery beats the ring-buffer's shared-batch indexed-read access pattern. The design's win is workload-shape-dependent: TPC-H's broad variety is the case for the design choice; ClickBench's narrow stress on adversarial shapes is the case for tuning $K$ and consumer access patterns when those shapes dominate.

\section{Production Considerations}
\label{sec:production}

Oxla is Redpanda's distributed analytical SQL engine. Ring-buffer shuffle has been the intra-process redistribution primitive in production for two years. We summarize where shuffle is invoked, how the ring-buffer parameters and thread counts are configured in production, how the design behaves under failure, and the implementation lessons that emerged at production scale.

\subsection{Where Shuffle Is Used}
Three operator families invoke ring-buffer shuffle in Oxla query plans. \emph{Hash joins} shuffle both build and probe sides on the join key, so matching rows converge on the same thread. \emph{Hash aggregation} pre-shuffles rows by the grouping key for high-cardinality \texttt{GROUP BY}, where local-then-merge would materialize too much intermediate state. \emph{Window functions} with \texttt{PARTITION BY} route rows so all rows sharing a partition key are evaluated on a single thread, allowing per-thread ordered window computation without cross-thread coordination during evaluation.

\subsection{Tuning K and Batch Size}
Ring capacity $K$ is chosen based on cache topology: $K{=}1$ on chiplet or multi-socket hardware, $K{=}2$ on single-coherence-domain architectures such as the Grace-class servers, consistent with the sweep in \autoref{sec:eval-ring-k}. The group capacity is set to $G{=}M$ (one slot per producer thread), so each producer contributes approximately one batch before the group fills, keeping the slot counter contended for at most $G$ \texttt{fetch\_add} operations per group.

Batch size is configured per query as a function of input cardinality and column types: smaller cardinalities use 8192 rows; large analytic columns (strings, JSON) use smaller row counts to keep total per-batch bytes within the working-set budget identified in \autoref{sec:eval-ring-k}.

\subsection{Thread Counts}
Oxla uses $M{=}N$ in all shuffle instances: every core runs one producer and one consumer, co-scheduled on the same physical core. This maximizes core utilization by ensuring that when a producer stalls (e.g., on a full ring), the consumer on the same core can run, and vice versa. The ring-buffer design does not require $M{=}N$, but the group capacity $G$ is set proportional to $M$, so asymmetric configurations change the dynamics. When $M > N$, groups fill faster (more producers contributing) while fewer consumers drain them, increasing backpressure frequency. When $M < N$, groups fill slowly and consumers frequently block waiting for new groups; the temporal convergence described in \autoref{sec:cache-locality} becomes stronger (more consumers converge on each group), but the reduced production rate may underutilize consumer cores. All experiments in \autoref{sec:evaluation} use $M{=}N$; characterizing asymmetric configurations is left to future work.

\subsection{Failure and Cancellation}
The ring-buffer shuffle is single-use and scoped to one query plan node; there is no in-shuffle recovery. All error and cancellation paths converge on a single \texttt{stop()} primitive, which the query coordinator invokes whenever any thread reports an error or the query is cancelled.

\emph{Producer fault during slot write.} A producer first atomically increments \texttt{writes\_started} to claim a slot, writes its batch, then increments \texttt{writes\_completed}. If it throws between these two atomics (e.g., from a downstream evaluation error), \texttt{writes\_completed} never reaches the group capacity $G$, the full flag is never set, and the group is never published. Producers waiting to publish into the now-stuck insertion buffer also hang. The faulted thread reports the error and the coordinator calls \texttt{stop()}; all blocked threads then exit through the cancellation path.

\emph{Consumer cancellation.} A cancelled consumer skips the per-batch decrement on \texttt{consumers\_left}, so the group's ring slot is never returned to the free pool; producers eventually fill the ring and block on backpressure. There is no in-flight detection of consumer cancellation---it is always handled by query-level teardown via \texttt{stop()}.

\emph{Stop and error propagation.} \texttt{stop()} sets a finish flag under the queue mutex and broadcasts on every condition variable. All blocked threads observe the flag and exit cleanly; partial in-flight data is dropped rather than flushed. Errors captured by any producer or consumer are stored as a status code on the queue and surfaced when other threads next call into the queue, avoiding cross-thread exception propagation. Consolidating recovery into a single \texttt{stop()} primitive rather than building per-mode handling keeps the hot path free of failure-detection logic, which would add cost for a behavior invoked at most once per query.

\subsection{Lessons Learned at Production Scale}
The ring-buffer design described in this paper is itself a response to production scaling failures. Oxla's prior shuffle implementation was the channel-based design we use as the baseline in \autoref{sec:eval-tpch} (one MPSC channel per output partition); it scaled acceptably up to roughly 32 cores but collapsed at 64 cores and beyond as per-channel mutex contention dominated on the customer workloads of the time. The ring-buffer redesign was driven specifically by this 64-core wall, and the implementation techniques below were necessary to make the new design hold its $O(1)$ synchronization rate at production core counts. None of these techniques were part of the initial design---each emerged from observed production behavior that smaller-scale microbenchmarks did not surface. We note that the reconstructed channel-based build in \autoref{sec:eval-tpch}, running on modern Neoverse V2 hardware, does not reproduce that historical collapse uniformly: the ring-buffer advantage at the query level is workload-shape-dependent, and the original production motivation reflects a particular point in the engine's history rather than a permanent ranking between the two designs.

\emph{Selective producer notification.} An earlier implementation woke a producer as soon as a single ring slot was freed. At high core count this caused context-switch storms when consumers ran slightly faster than producers, with throughput collapsing under scheduler overhead. Notifying only when ring occupancy dropped below half capacity restored throughput.

\emph{Pre-allocated replacement groups.} The first implementation allocated each new group under the queue mutex. At 64+ producers this extended the critical section and showed up as a flat mutex-contention plateau in profiles. Donating a producer-owned replacement via pointer swap moved the allocation off the critical path.

\emph{Per-producer buffer references.} An early design used a shared atomic pointer to the current insertion buffer. The atomic refcount became a cross-core cache-line hotspot; the per-producer container with private mutexes eliminated this contention.

\section{Conclusion}
\label{sec:conclusion}
Intra-process shuffle---how a query engine repartitions data across parallel execution threads within a single server---is a critical primitive whose design space has received surprisingly little dedicated study. In this paper, we framed the problem in terms of a fundamental tension between memory overhead and synchronization contention, and systematically compared three designs that span the resulting tradeoff space: batch partitioning, channel-based streaming, and ring-buffer streaming. Our ring-buffer approach achieves amortized $O(1)$ synchronization cost per batch and $O(M)$ memory through atomic slot acquisition and fixed-size batch groups. Several implementation techniques---per-producer buffer references, pre-allocated replacement groups, and selective notification---proved essential to realizing these properties at high core counts.

Our evaluation reveals that the relative advantage of each design depends critically on cache topology and core count. On the 72-core unified LLC GraceHopper system (Arm Neoverse V2), ring-buffer streaming outperforms channel streaming by up to 44\% and batch partitioning by up to 79\%. At 192 cores on a dual-socket Graviton4, the advantage over channel grows to over 100\% and over 300\% versus batch partitioning---both because higher core counts amplify channel contention, and because the Graviton4's higher aggregate memory bandwidth keeps the workload synchronization-bound across all tested batch sizes, whereas on Grace and EPYC memory bandwidth saturates earlier and compresses the gap. However, on chiplet architectures with many partitioned L3 caches (AMD EPYC, 12 CCDs), the ring buffer's shared atomic counter becomes a cross-CCD bottleneck, and channel-based streaming remains competitive at medium and large batch sizes. The gap narrows as batch sizes grow and the workload shifts from synchronization-bound to memory-bandwidth-bound, though at 192 cores the ring buffer retains a substantial lead even at the largest batch sizes due to growing congestion of the channel approach.

Our channel-based baseline uses multi-producer single-consumer (MPSC) channels; an alternative using $M \times N$ dedicated single-producer single-consumer (SPSC) channels could reduce per-channel contention at the cost of higher memory, and is a natural baseline for future comparison. The end-to-end results in \autoref{sec:e2e} show that the ring-buffer's wins and losses are predictable from query plan shape (specifically, the downstream operator's access patterns), suggesting that a query optimizer could select the shuffle implementation per pipeline stage at plan time. Another promising direction for future work is a NUMA-aware ring-buffer variant that uses per-CCD batch groups with local atomic counters, avoiding cross-die traffic on the critical acquisition path while preserving the ring buffer's streaming and memory-efficiency advantages on chiplet architectures.

\subsection{Acknowledgments}

The authors would like to thank the members of the Oxla team who have helped build and maintain our query engine over the years: Wojciech Chlapek, Grzegorz Dudek, Mikołaj Gagatek, Jacek Gałązka, Krzysztof Grabski, Mateusz Grzonka, Marcin Grzebieluch, 
  Michał Korbel, Witold Kozłowski, Franciszek Kucza, Kajetan Litwinowicz, Michal Maślanka, Jan Mikuła, Tomasz Nowak, Wojciech Oziębły, Wojciech Padło,
  Konrad Piotrowski, Mariusz Rokicki, Jacek Seliga, Krzysztor Smogów, Paweł Sobótka, Vagelis Sofikitis, Paweł Szczur,       
  Szczepan Szpilczyński, Przemek Zglinicki.


\balance
\bibliographystyle{ACM-Reference-Format}
\bibliography{sample}

\end{document}